\documentclass[fleqn,usenatbib]{mnras}

\usepackage{newtxtext,newtxmath}

\usepackage[T1]{fontenc}

\DeclareRobustCommand{\VAN}[3]{#2}
\let\VANthebibliography\thebibliography
\def\thebibliography{\DeclareRobustCommand{\VAN}[3]{##3}\VANthebibliography}

\usepackage{graphicx}	
\usepackage{amsmath}	

\title[A Quadruple Excess in Wide Binaries]{A Quadruple Excess in Wide Binary Systems: Evidence for Correlated Binary Formation}

\author[Bashi, Clarke, Belokurov]{
Dolev Bashi$^{1}$\thanks{E-mail: db975@cam.ac.uk},\,
Cathie J. Clarke$^{2}$\ and  
Vasily Belokurov$^{2}$\\
$^{1}$Astrophysics Group, Cavendish Laboratory, University of Cambridge, JJ Thomson Avenue, Cambridge CB3 0HE, UK\\
$^{2}$Institute of Astronomy, University of Cambridge, Madingley Road, Cambridge CB3 0HA, UK\\
}

\date{Accepted XXX. Received YYY; in original form ZZZ}

\pubyear{\the\year{}}

\begin{document}
\label{firstpage}
\pagerange{\pageref{firstpage}--\pageref{lastpage}}
\maketitle

\begin{abstract}
Understanding the multiplicity of stellar systems and the correlations between their hierarchical components provides crucial insights into star formation processes. If binary companions form independently in each component of a wide binary (WB), the fraction of quadruple systems, i.e., 2+2 configurations where both components are themselves close binaries (CBs), should equal the product of individual CB fractions. Using \textit{Gaia} DR3 radial velocity spectroscopy (RVS) data for WB systems, we measure the CB fraction $p$ and quadruple fraction $P_{2+2}$, suggesting an enhancement factor $\kappa = P_{2+2}/p^2 = 2.34_{-0.11}^{+0.12}$, significantly exceeding unity expected under a statistical model of independence. We confirm the significance of this excess by performing two sets of tests: (1) shuffling WB pairings while preserving the overall $\Delta G$ distribution shows no significant enhancement, ruling out selection effects; (2) simulations preserving the spectral type (temperature-dependent) CB fraction also yield the same null excess. When examined as a function of WB separation, the enhancement remains strong at separations $\leq 5\,000$ AU, but shows a decline towards unity at the widest separations ($\geq 10\,000$ AU). An independent proper motion anomaly (PMa) consistency check confirms the enhancement, suggesting a similar value. We further find that the enhancement declines with increasing peculiar velocity, suggesting that dynamical processing in older or dynamically hotter populations may transform 2+2 quadruples into triples over time. Our results provide strong evidence for correlated binary formation processes operating in WB systems.

\end{abstract}

\begin{keywords}
binaries: close -- binaries: general -- binaries: spectroscopic -- stars: formation -- methods: statistical
\end{keywords}



\section{Introduction}
\label{sec:intro}
Binary and multiple star systems are ubiquitous in the Galaxy, with the binary
fraction among solar-type stars estimated to be 40--60\%
\citep{Raghavan10,MoeDitefano17,Offner23}. Understanding the formation and evolution of multiple stellar systems provides crucial constraints on star formation theories and on the initial conditions of stellar populations.

Hierarchical multiple systems, in which one or both components of a wide binary (WB) host close companions, offer a particularly powerful diagnostic of the
formation process. If close binaries (CB) form independently in each component of a WB, the fraction of systems in which both components are binaries, i.e., 2+2 quadruple systems, should simply equal the product of the individual CB fractions. Deviations from this expectation may signal correlated
formation processes, environmental effects, or preferential survival of
specific configurations.

Early observational evidence for such deviations was reported by
\citet{TokovininSmekhov02}, who noted a deficit of single stars in triple systems and an excess of 2+2 quadruple systems. This result was later confirmed by \citet{Tokovinin14}, who found that among solar-type field stars the 2+2 quadruple fraction exceeded expectations from independent binary formation by a factor of $\sim2.6$. However, these studies were based on relatively small and heterogeneous samples, leaving open the possibility of residual selection biases.

More recently, \citet{Fezenko22} used eclipsing binary candidates in wide systems from \textit{Gaia} EDR3 to investigate the quadruple fraction, finding a $\sim 7$ times higher occurrence rate of 2 + 2 quadruples compared to random pairings of field stars. That analysis was nevertheless limited by the low number of systems analysed in their work (eight eclipsing 2+2 quadruples), given the selection effects associated with eclipse geometry.

From a theoretical perspective, WBs and CBs are generally thought to arise through distinct physical processes operating on different spatial and temporal scales. WBs are commonly associated with the fragmentation of molecular cores or filaments and with dynamical processes during the early phases of cluster evolution, making their formation and survival sensitive to the natal environment \citep[e.g.,][]{Kouwenhoven10,MoeckelClarke11, Lee17, RoznerPerets23}. In contrast, CBs are typically linked to mechanisms acting at smaller scales, such as disc fragmentation and subsequent orbital evolution \citep[e.g.,][]{Bate95, Clarke09, TokovininMoe20}. In hierarchical systems, these processes may occur sequentially, with close subsystems forming within the components of an already established wide pair. If so, correlations between the multiplicity properties of the two components are plausible, motivating an empirical test of whether the occurrence of 2+2 quadruple systems is consistent with independent binary formation or instead exhibits a statistically significant enhancement.

The \textit{Gaia} mission \citep{GaiaCollaboration2016} has revolutionised the
study of stellar multiplicity \citep{Arenou23}. Large and clean samples of wide
binaries can now be identified with high confidence using precise parallax and
proper motion information \citep{El-Badry21}. Crucially, \textit{Gaia}'s Radial
Velocity Spectrometer \citep[RVS;][]{Cropper18,Katz23} provides complementary
spectroscopic information that enables a robust census of CB companions through RV variability
\citep{Bashi24,BashiBelokurov25}.

In this paper, we exploit the availability of \textit{Gaia} RVS data for a
substantial fraction of WB systems to perform a comprehensive and
comparatively unbiased census of hierarchical multiplicity. We aim to test
whether the observed 2+2 quadruple fraction is consistent with independent
binary formation or instead exhibits a statistically significant enhancement,
and to explore potential physical implications.

The paper is organised as follows. Section~\ref{sec:sample} describes our sample
selection. Section~\ref{sec:methods} outlines our methodology for measuring close
binary fractions and the excess of 2+2 systems. Section~\ref{sec:results}
presents our main results and validation tests. Section~\ref{sec:discussion}
discusses potential formation pathways and the peculiar velocity trend.
Section~\ref{sec:conclusions} summarises our conclusions.

\section{Sample Selection}
\label{sec:sample}

\subsection{Wide Binary Catalogue}

Our analysis is based on the WB catalogue of \citet{El-Badry21}, which provides a high-purity sample of WB systems identified using \textit{Gaia} EDR3 astrometry. The catalogue employs a probabilistic approach, computing the probability that a pair of stars is a true binary based on their positions, parallaxes, proper motions, and the local stellar density. We use pairs with a chance alignment probability $R_{\rm chance} < 10^{-3}$, ensuring high confidence that our sample consists of gravitationally bound systems.


\subsection{Gaia DR3 RVS Data}

We cross-match the WB catalogue with \textit{Gaia} DR3 to obtain RVS measurements for both components. Following \citet{BashiBelokurov25}, we require reliable RV measurements (\texttt{rv\_method\_used} = 1)~corresponding to bright stars ($G_{\rm RVS} < 12$), positive parallaxes, and available peak-to-peak RV amplitudes (\texttt{rv\_amplitude\_robust}). The \texttt{rv\_amplitude\_robust} field quantifies the peak-to-peak variability of the RV time series, defined as the difference between the maximum and minimum robust RV measurements after outlier rejection \citep{Katz23}. Throughout this work, we denote this quantity as $\mathrm{RV_{pp}}$. To mitigate the large radial-velocity uncertainties in hot stars \citep{Blomme23}, we restrict our sample to cool stars with
$3900< \texttt{rv\_template\_teff} < 6900 ~\mathrm{K}$,
and require more than eight RV visibility periods (\texttt{rv\_visibility\_periods\_used} $> 8$) to ensure adequate phase coverage.

Requiring both components of each WB to satisfy these criteria, yields a sample of $12\,128$ WB pairs with high-quality RVS data. 

To ensure a homogeneous stellar population and minimise the effects of stellar evolution on binary properties, we restrict our sample to main-sequence (MS) dwarfs. We define an MS selection region in the colour-magnitude diagram (CMD) using \textit{Gaia} photometry ($G$, $G_{\rm BP}$, $G_{\rm RP}$) and parallax-based absolute magnitudes. We apply a polygon cut in the $(G_{\rm BP} - G_{\rm RP}, M_G)$ plane with vertices at [(0.25, 4.0), (0.25, 2.0), (0.95, 2.0), (1.0, 4.0), (2.2, 7.0), (2.2, 9.2), (0.25, 4.0)], approximately enclosing F, G, and K dwarfs. 
We require that the \emph{primary component} (defined as the brighter star in $G$) falls within this MS region. After applying these cuts, our final sample consists of $9\,411$ WB pairs with RVS measurements for both components\footnote{The full list of sources used in this analysis is provided in the online supplementary material.}. 

\section{Methods}
\label{sec:methods}

\subsection{RVS Binary Fraction}

We employ a two-stage hierarchical Bayesian framework to measure binary fractions in a sample of \textit{Gaia} RVS sources. \emph{Stage-1} determines the CB fraction for individual components, while \emph{Stage-2} measures the fraction of pairs where both components are CBs, i.e., the 2+2 quadruple fraction.

\subsubsection{Stage-1: Component-Level Fits}
Following \citet{Bashi24,BashiTokovinin24,BashiBelokurov25}, we model the distribution of 
$x = \log_{10}({\rm RV_{pp}}/2)$ as a function of RVS magnitude $G_{\rm RVS}$ as a mixture of single stars and RV-jittered (CB candidates)
\begin{equation}
f(x \mid G_{\rm RVS}; \theta) =
(1-p)\,\mathcal{N}_{\rm s}(x \mid \mu_{\rm s}, \sigma_{\rm s}^2)
+ p\,\mathcal{N}_{\rm b}(x \mid \mu_{\rm b}, \sigma_{\rm b}^2),
\end{equation}
where  
$\theta = (p, a, b, G_{\rm 0}, d, \sigma_{\rm s}, \sigma_{\rm b})$ denotes the full set of model parameters.
Here, $p$ denotes the fraction of systems that are unresolved CBs, $a,b,G_{\rm 0}$ define the magnitude dependence of the single-star RV scatter, $d$ quantifies the additional RV variability induced by orbital motion in binaries,
and $\sigma_{\rm s}$ and $\sigma_{\rm b}$ describe the intrinsic dispersions of the single-star and binary components, respectively.
Both components are modelled as right-truncated Gaussians to reflect the physical \textit{Gaia} limit on measurable RV amplitudes \citep{Katz23}.

Using this approach, the single-star mean is parametrised as
\begin{equation}
\mu_{\rm s}(G_{\rm RVS}) = a + e^{b(G_{\rm RVS} - G_{\rm 0})},
\end{equation}
capturing the increase in RV noise towards fainter magnitudes.
The binary component is offset relative to the single-star distribution by the additional variability term $d$,
\begin{equation}
\mu_{\rm b}(G_{\rm RVS}) =
\log_{10}\!\left[\sqrt{10^{2\mu_{\rm s}(G_{\rm RVS})} + d^2}\right],
\end{equation}
which accounts for unresolved orbital RV motion.

We fit this model using Markov Chain Monte Carlo (MCMC) sampling with \texttt{emcee} \citep{ForemanMackey2013},
adopting the same priors and implementation details as in \citet{BashiBelokurov25} and listed in Table~\ref{tab:stage1_priors}. 

A qualitative indication of the orbital periods probed by this selection is given by the Gaia-RVS forward modelling of \citet{BashiBelokurov25}, which shows that the method is mainly sensitive to short- and intermediate-period binaries, with declining completeness toward longer periods ($P > 1\,000 $ days).

We estimate the CB fractions of the primary and secondary components of WBs, denoted $p_{\rm 1}$ and $p_{\rm 2}$, respectively.
To obtain a global CB fraction $p$, we construct a \emph{pooled} sample by combining both components from all valid WB pairs and fit the Stage-1 model to this combined dataset.

\subsubsection{Stage-2: Joint Binary Fraction}

To measure the fraction $P_{2+2}$ of wide pairs where \emph{both} components are CBs, we extend the mixture model to the joint distribution. Given Stage-1 fits for each component, providing magnitude-dependent single and binary distributions, we model the joint sample as a mixture of four populations:
\begin{align}
P(x_1, x_2) &= q_{00} \, \mathcal{N}_{\rm s,1}(x_1) \mathcal{N}_{\rm s,2}(x_2) \nonumber \\
&\quad + q_{10} \, \mathcal{N}_{\rm b,1}(x_1) \mathcal{N}_{\rm s,2}(x_2) \nonumber \\
&\quad + q_{01} \, \mathcal{N}_{\rm s,1}(x_1) \mathcal{N}_{\rm b,2}(x_2) \nonumber \\
&\quad + q_{11} \, \mathcal{N}_{\rm b,1}(x_1) \mathcal{N}_{\rm b,2}(x_2),
\end{align}
where $q_{ij}$ is the fraction in state $(i,j)$ with $i,j \in \{0,1\}$ indicating single (0) or binary (1). 
By construction, the marginal fractions satisfy $q_{10}+q_{11}=p_1$ and $q_{01}+q_{11}=p_2$,
where $p_1$ and $p_2$ are the CB fractions 
as determined from the Stage-1 fits.
Under this parametrisation, the quadruple fraction is simply $q_{11} \equiv P_{2+2}$.
We infer $P_{2+2}$ using MCMC sampling, holding the marginal CB fractions $p_1$ and $p_2$ fixed to their Stage-1 values.
We adopt a uniform prior on $P_{2+2}$ over its feasible interval,
\begin{equation}
P_{2+2} \sim \mathcal{U}\!\left(\max\!\left[0,\,p_1+p_2-1\right],\ \min[p_1,p_2]\right),
\end{equation}
which enforces $q_{ij}\ge 0$ and $\sum_{ij} q_{ij}=1$ given fixed marginals $p_1$ and $p_2$.

Using this model, we define the excess factor as
\begin{equation}
\kappa \equiv \frac{P_{2+2}}{p^2}.
\end{equation}
Values of $\kappa = 1$ indicate independent binary formation, while $\kappa > 1$ suggests correlated formation. 

We note that, because our sample is magnitude limited, the absolute value of the inferred CB fraction $p$ is not expected to represent the intrinsic, volume-complete binary fraction. In particular, brighter systems, including binaries at larger distances or with more luminous components, are preferentially selected. However, this limitation does not affect the derivation of the excess factor $\kappa$, which depends only on the \emph{relative} occurrence of CBs in the two components of the same wide pair. Since both stars in each WB are subject to identical selection criteria, and the marginal CB fractions $p_1$ and $p_2$ are measured consistently from the same magnitude-limited parent sample, any global incompleteness largely cancels in the ratio. As a result, $\kappa$ provides a robust measure of departures from independent binary formation, even if the absolute normalisation of $p$ is biased.

As an independent astrometric consistency check, we also compute a proper motion anomaly (PMa)–based binary classification using \textit{Gaia} DR2–DR3 proper motion differences. The PMa methodology and resulting $\kappa_{\rm PMa}$ measurement are described in Appendix~\ref{appendix:PMa}.


\section{Results}
\label{sec:results}

\subsection{Close Binary Fraction and Enhancement}

\begin{figure}
\includegraphics[width=\columnwidth]{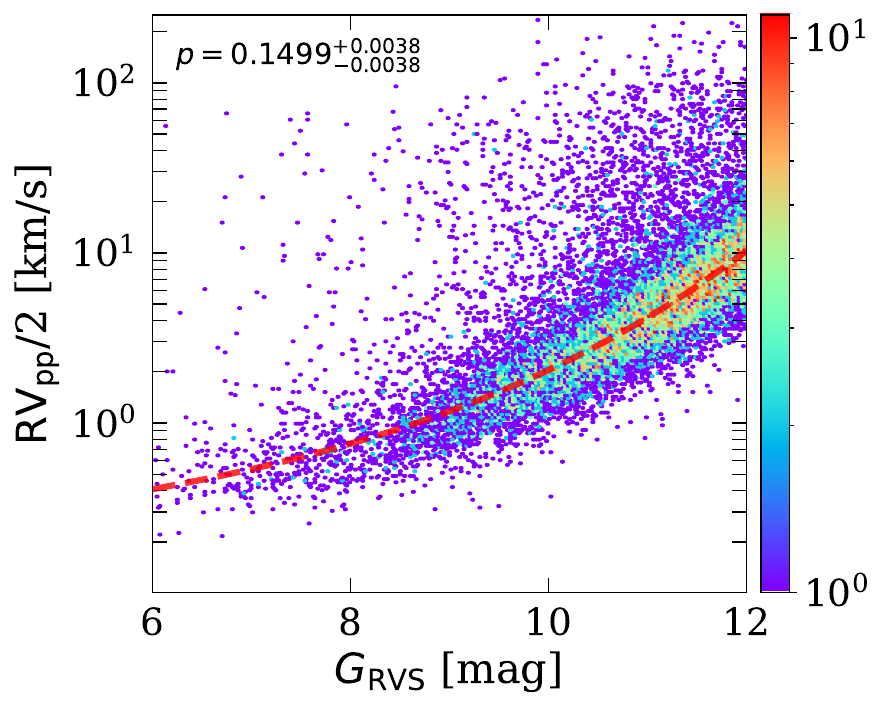}
\caption{
RV semi-amplitude as a function of $G_{\rm RVS}$ for the pooled WB sample of $18\,822$ stars. 
The density map shows the distribution of $\mathrm{RV_{pp}}/2$ measurements, while the red dashed curve indicates the best-fitting single-star model inferred from the Stage-1 MCMC, which captures the magnitude-dependent RV noise floor. 
Systems with RV variability significantly above this locus are interpreted as hosting CBs and form the basis for the Stage-2 joint analysis.
}
\label{fig:rv_grvs}
\end{figure}

Figure~\ref{fig:rv_grvs} illustrates the distribution of RV semi-amplitudes as a function of $G_{\rm RVS}$ for the pooled sample of WB components, shown as an example of the Stage-1 modelling procedure. The red dashed curve shows the best-fitting single-star locus inferred from the Stage-1 MCMC, which captures the magnitude dependence of the RV noise floor. This model provides the baseline against which excess RV variability, attributed to CBs, is identified. The adopted priors and the resulting posterior constraints for all Stage-1 model parameters are summarised in Table~\ref{tab:stage1_priors} for the pooled sample and for each component separately. The lower inferred CB fraction for Comp. 2 is consistent with its lower typical stellar mass, while the larger value of $d$ likely reflects a selection effect: because Comp. 2 is fainter, only binaries with larger absolute RV variability are cleanly separated from the single-star locus.

The Stage-1 pooled MCMC fit yields a CB fraction $p = 0.150 \pm 0.004$. The Stage-2 joint fit gives the quadruple fraction $P_{2+2} = 0.053 \pm 0.003$, corresponding to an enhancement factor 
\begin{equation}
\kappa = \frac{P_{2+2}}{p^2} = 2.34_{-0.11}^{+0.12}.
\end{equation}
Our result provides strong evidence that quadruple systems are more than twice as common as expected from independent binary formation.

An independent PMa-based analysis, sensitive to longer-period companions, yields a consistent enhancement (Appendix~\ref{appendix:PMa}).

\begin{table}
\centering
\caption{Stage-1 priors and posterior constraints for the mixture model parameters, shown for the pooled sample and for components 1 and 2 separately.}
\label{tab:stage1_priors}
{\fontsize{6.5}{10}\selectfont
\begin{tabular}{lcccc}
\hline
Parameter & Prior & Pooled & Comp.\ 1 & Comp.\ 2 \\
\hline
$p$ & $\log \mathcal{U}(0.001,1)$ & $0.150_{-0.004}^{+0.004}$ & $0.167_{-0.005}^{+0.005}$ & $0.127_{-0.005}^{+0.005}$ \\
$a$ & $\mathcal{U}(-5,0.5)$ & $-0.822_{-0.029}^{+0.027}$ & $-0.801_{-0.036}^{+0.034}$ & $-0.963_{-0.052}^{+0.047}$ \\
$b$ & $\log \mathcal{U}(10^{-3},1)$ & $0.241_{-0.006}^{+0.005}$ & $0.245_{-0.007}^{+0.008}$ & $0.223_{-0.007}^{+0.008}$ \\
$G_{\rm 0}$ & $\mathcal{U}(0,15)$ & $9.496_{-0.115}^{+0.117}$ & $9.563_{-0.145}^{+0.143}$ & $8.961_{-0.216}^{+0.213}$ \\
$d$ & $\log \mathcal{U}(10^{-0.4},10^{1.6})$ & $12.812_{-0.478}^{+0.436}$ & $10.492_{-0.493}^{+0.500}$ & $17.299_{-0.886}^{+0.930}$ \\
$\sigma_s$ & $\log \mathcal{U}(0.001,10)$ & $0.154_{-0.001}^{+0.001}$ & $0.152_{-0.001}^{+0.002}$ & $0.156_{-0.001}^{+0.001}$ \\
$\sigma_b$ & $\log \mathcal{U}(0.01,10)$ & $0.455_{-0.008}^{+0.007}$ & $0.470_{-0.009}^{+0.010}$ & $0.436_{-0.012}^{+0.010}$ \\
\hline
\end{tabular}}
\end{table}

\subsection{Validation Tests}

To ensure that the observed enhancement in the 2+2 quadruple fraction does not arise from selection effects or hidden correlations in the WB sample, we perform two independent validation experiments. These tests are designed to isolate, respectively, biases associated with the WB pairing itself and biases related to the underlying CB occurrence as a function of stellar properties.

\subsubsection{Wide binary shuffling}

Given that both components of a wide binary are required to have a measured $\mathrm{RV_{pp}}$, the sample is effectively magnitude limited, which introduces a bias in the distribution of component magnitudes and favours twin WBs, i.e., near-equal brightness pairs \citep{El-Badry19}. To test whether this selection effect could artificially
produce an apparent enhancement of 2+2 systems, we construct synthetic WB catalogues that preserve the relevant observational distributions while removing any
physical correlation between the components.

We divide the sample into distance bins with edges at
$(0,\ 100,\ 150,\ 200,\ 250,\ 300,\ 400,\ 1\,000)\ \mathrm{pc}$, based on parallax and treat each bin independently. Within each distance bin, we measure the observed joint distribution
of primary and secondary magnitudes, $(G_1, G_2)$, enforcing the physical constraint
$G_2 \ge G_1$. Synthetic pairs are then assembled by sampling target $(G_1, G_2)$ values from this
two-dimensional distribution and pairing stars drawn from the pooled set of components
within the same distance bin. For each target pair, two distinct stars whose magnitudes
most closely match the sampled $(G_1, G_2)$ values are selected (without reuse), with
the brighter star assigned as the primary. This procedure preserves the distance distribution as well
as the full joint $(G_1, G_2)$ and $\Delta G$ distributions, while explicitly destroying any physical correlation between the two components beyond those induced by the preserved observational distributions.

We generate $10^4$ such realisations and compute $\kappa$ for each using the Stage-1 and Stage-2 MCMC framework. Because the reshuffling changes which stars enter the primary and secondary subsamples, the component-level CB fractions 
($p_1$ and $p_2$) are re-estimated in each realisation while the pooled fraction $p$ remains unchanged, since the underlying set of stars is fixed. In all cases, the null distribution is constructed under identical observational selection, differing only by the removal of any physical correlation between WB components.

\begin{figure}
\includegraphics[width=\columnwidth]{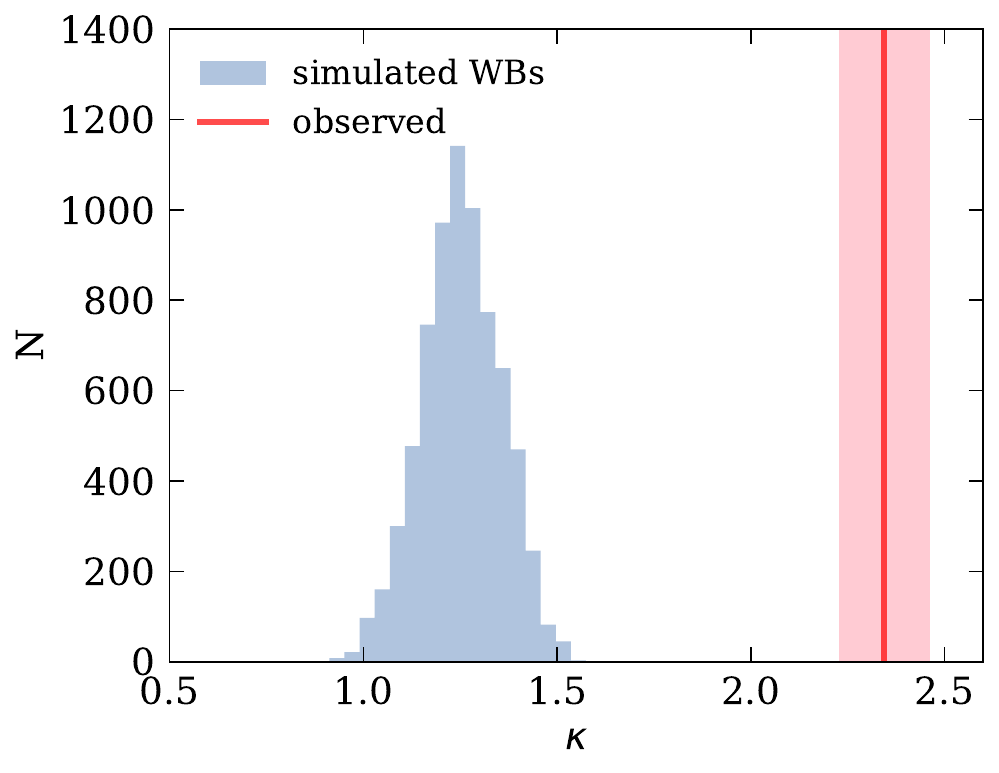}
\caption{Distribution of $\kappa = P_{2+2}/p^2$ obtained from $10^4$ realisations in which both components of WBs are randomly reassembled within distance bins while preserving the observed joint $(G_1, G_2)$ distribution. For each realisation, $\kappa$ is computed using the full Stage-1 and Stage-2 MCMC framework. 
While the pooled Stage-1 fit (and thus $p$) is unchanged, the component-level fractions $p_1$ and $p_2$ are re-estimated for each shuffled catalogue, and Stage-2 is re-run to infer $P_{2+2}$. The vertical red line and shaded regions mark the observed value of $\kappa$ and corresponding uncertainties from the real sample.}
\label{fig:Wide_sim}
\end{figure}

Figure~\ref{fig:Wide_sim} shows the results, where the shuffled samples yield a median $\langle \kappa_{\rm sim} \rangle ^{\rm WB} = 1.25$, consistent with independent pairing. We note that the slight upward offset of $\langle \kappa_{\rm sim} \rangle ^{\rm WB} $ above unity arises naturally from a distance-dependent variation of the component CB fraction within our magnitude-limited sample. As reshuffling is performed within distance bins, stars with similar intrinsic CB probabilities are preferentially paired, which mildly enhances both 2+2 and 1+1 configurations even under complete independence. None of the $10^4$ realisations exceeded the observed $\kappa = 2.34$, 
implying a tail probability $p < 10^{-4}$ under the null hypothesis. This confirms that the enhancement is highly significant and not driven by 
selection effects related to distance or magnitude differences in the WB pairing.

\subsubsection{Close binary shuffling}

As a second validation, we test whether the observed enhancement could arise from a correlation between the WB pairing selection, which favours twin systems, and the known dependence of the CB fraction on spectral type \citep[e.g.,][]{MoeDitefano17, Offner23}. To minimise the impact of metallicity-dependent multiplicity trends, we restrict the analysis to a subset of $4\,219$ systems with metallicities in the range -0.2 < [Fe/H] < 0.1 dex, based on the values reported by \citet{Andrae23}. This selection avoids the well-established anti-correlation between CB fraction and metallicity \citep[e.g.,][]{Moe19, Price-Whelan20, Bashi24, BashiBelokurov25}.

Using effective-temperature–binned measurements that span $T_{\rm eff} = 4000$--$7000~\mathrm{K}$, we assign each stellar component a temperature-dependent CB probability $p(T_{\rm eff})$ according to an empirically fitted relation, $p(T_{\rm eff}) = m\,T_{\rm eff} + n$, with $m = 7.88\times10^{-5}~\mathrm{K^{-1}}$ and $n = -0.291$. In this temperature range, the observed dependence of the CB fraction on $T_{\rm eff}$ is well approximated by a linear relation. We then perform Monte Carlo simulations in which the two components of each WB are independently flagged as CBs according to their assigned $p(T_{\rm eff})$. This procedure preserves the observed temperature distribution of the sample while explicitly removing any correlation between WB pairing and CB occurrence, allowing us to assess whether the measured enhancement can be reproduced purely by temperature-dependent CB statistics.

\begin{figure}
\includegraphics[width=\columnwidth]{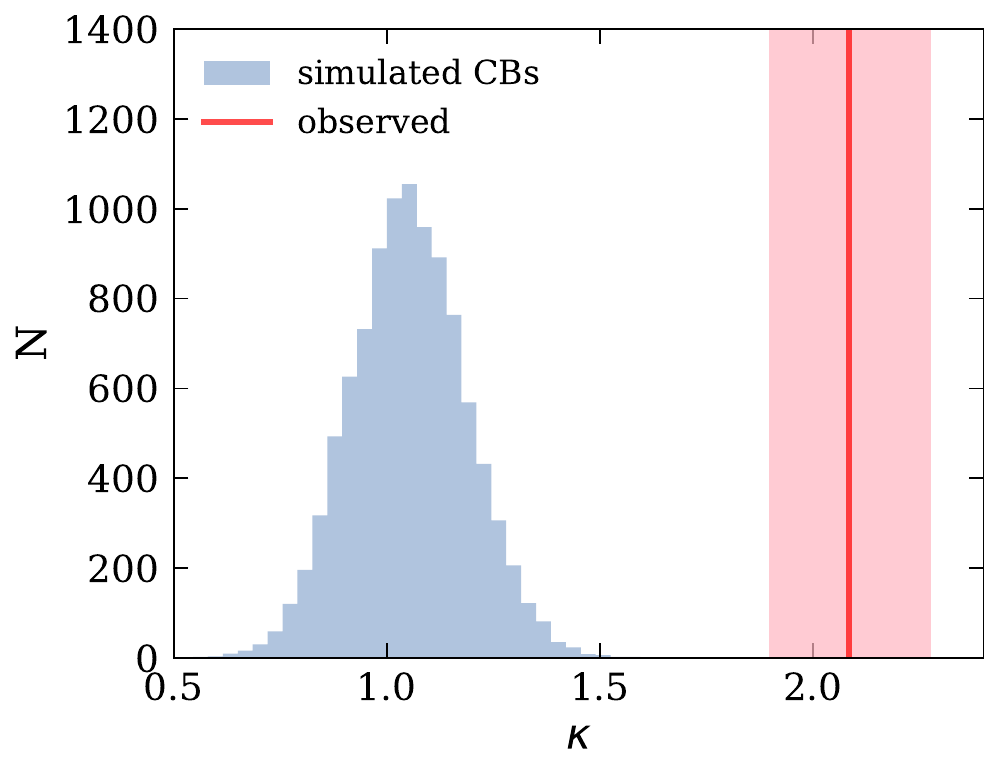}
\caption{Distribution of $\kappa = P_{2+2}/p^2$ obtained from $10^4$ realisations in which WB components are independently flagged as CBs using a temperature-dependent probability $p(T_{\rm eff})$ defined based on a subset of $4\,219$ systems with metallicities in the range -0.2 < [Fe/H] < 0.1 dex. The vertical red line and shaded regions mark the observed value of $\kappa$ and corresponding uncertainties from the real sample.}
\label{fig:close_sim}
\end{figure}

Figure~\ref{fig:close_sim} shows the resulting $\kappa$ distribution from $10^4$ simulations. The simulations yield a median $\langle \kappa_{\rm sim} \rangle^{\rm CB} = 1.05$, 
consistent with the independence hypothesis. None of the $10^4$ realisations exceeded the observed value $\kappa = 2.09$, implying a tail probability $p < 10^{-4}$ under the null hypothesis. The observed excess in our metal-limited sample therefore lies far in the extreme tail of the simulated distribution, ruling out spectral-type-dependent binary fractions as the source of the enhancement.

\subsection{Separation Dependence}

If the excess of 2+2 systems reflects either a common formation pathway or subsequent dynamical coupling between the two subsystems, it may depend on the outer separation of the WB. The projected physical separation, therefore, provides a natural observable proxy for how strongly the two inner systems are connected: tighter WBs are expected to retain a stronger imprint of shared formation conditions and, potentially, stronger secular coupling, whereas at sufficiently large separations the components should increasingly behave as independently paired systems.

\begin{figure*}
\includegraphics[width=12cm]{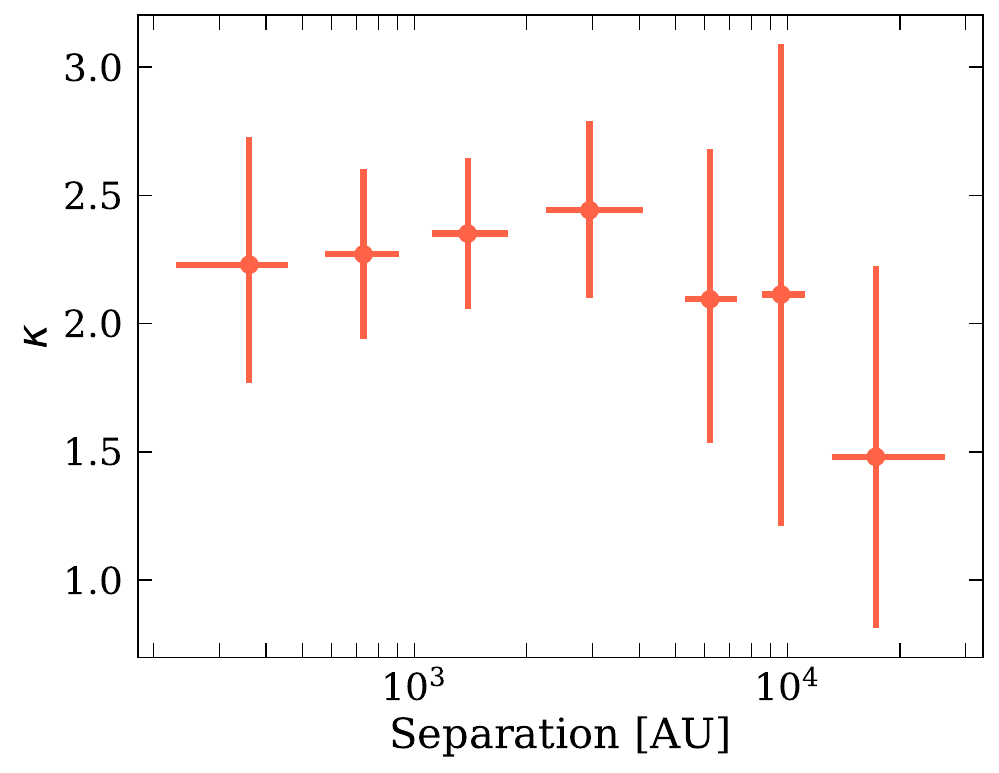}
\caption{Enhancement factor $\kappa = P_{2+2}/p^2$ as a function of WB separation. Points show measurements in separation bins with $1\sigma$ uncertainties from MCMC posteriors. The enhancement is strong ($\kappa \approx 2$--3) at separations $\lesssim 5\,000$ AU. At wider separations, large uncertainties prevent strong conclusions, though there is a suggestion of a decreasing enhancement.}
\label{fig:kappa_sep}
\end{figure*}

To investigate whether the enhancement varies with WB separation, we divide the sample into projected physical separation bins and compute $\kappa$ in each bin. We adopt logarithmically spaced bins motivated by both the separation distribution of the sample and the need to maintain sufficient statistics in each bin, with edges at
$(10,\ 500,\ 1\,000,\ 2\,000,\ 5\,000,\ 8\,000,\ 12\,000,\ \infty)\ \mathrm{AU}$.
Figure~\ref{fig:kappa_sep} shows the results.

At separations $\lesssim 5\,000$ AU, we find $\kappa \approx 2-3$, consistent with the global value. At the widest separations ($\gtrsim 10\,000$ AU), the measurements are consistent with unity within uncertainties, although the error bars are large. This trend is qualitatively consistent with the expectation that physical coupling weakens with increasing separation, and the WB population asymptotically approaches the regime of independent stellar pairing.

\section{Discussion}
\label{sec:discussion}

The observed enhancement $\kappa = 2.34_{-0.11}^{+0.12}$ indicates that WB systems with both components hosting CBs are more than twice as common as expected from random pairing. 

Our result is in excellent agreement with previous findings. \citet{Tokovinin14} analysed 1,747 outer binaries among solar-type field dwarfs and found 41 systems with 2+2 configurations. Under independent formation, the observed close binary fractions would predict only 16 such systems, yielding an enhancement factor of $41/16 \approx 2.6$, consistent with our measurement.

\citet{Fezenko22} identified $8$ eclipsing 2+2 quadruples among 
wide systems in a parent sample of $131\,245$ pairs, including $1\,282$ eclipsing binaries. Interpreting the eclipsing binary fraction as 
$1282/(2 \times 131245)$ and applying our definition 
$\kappa = P_{2+2}/p^2$, we obtain 
$\kappa \approx (8/131245) / (1282/(2 \times 131245))^2 = 2.55$ 
remarkably consistent with both our result and that of \citet{Tokovinin14}.

The agreement across multiple independent studies spanning  from small ground-based spectroscopic samples \citep{TokovininSmekhov02, Tokovinin14}, to eclipse-based surveys \citep{Fezenko22}, to our large RVS-based analysis and an independent PMa consistency check (Appendix~\ref{appendix:PMa}), establishes the enhanced 2+2 quadruple fraction as a robust empirical result requiring explanation by star formation theory. 

A related question, which we leave for future work, is whether the pooled CB fraction among WB components differs from that of a carefully matched field-star control sample. Addressing this would require a dedicated control sample with matched stellar and observational properties, together with additional checks for selection effects and contamination.

\subsection{Formation Mechanisms}

One possibility is that CBs in each component form through correlated fragmentation in the same molecular cloud core \citep{Lee19, TokovininMoe20}. If conditions promoting binary formation (e.g., angular momentum, turbulence, magnetic fields) are coherent over the $\sim 1\,000$ AU scale of WBs, both components may preferentially form with close companions, naturally producing $\kappa > 1$.

Alternatively, dynamical interactions during star formation may play a key role.  \citet{Tokovinin26} recently discussed how encounters between binaries, disruption of triples, or misaligned gas capture could reshape hierarchical systems and potentially explain the enhanced quadruple fraction. Encounters between two binaries can produce misaligned triples, while gas captured by inner subsystems may trigger fast migration \citep{Bate2019,Offner23}, a potentially dominant channel for CB formation in hierarchical systems.

Another possible contribution comes from the fact that if WBs are assembled from already-formed subsystems, then CB subsystems may be preferentially incorporated into wide pairs because of their larger total masses, potentially boosting the formation of 2+2 hierarchies relative to random pairing \citep{RoznerPerets23}.

Quantitative predictions from detailed simulations remain limited. Future hydrodynamic simulations of hierarchical fragmentation in turbulent cores, including magnetic fields, will be required to test these scenarios.

\subsection{Kinematic and Orbital Signatures of 2+2 Systems}

To place the properties of the 2+2 systems in context, we compare them to a matched control sample of ordinary WBs selected to have similar separations and component properties.

We identify $140$ 2+2 quadruple candidates\footnote{The full list of quadruple candidates is provided in the online supplementary material.} by requiring that both WB components show RV variability significantly above the magnitude-dependent single-star locus in $x=\log_{10}(\mathrm{RV_{pp}}/2)$. Specifically, for each component $i\in\{1,2\}$ we compute the residual
\begin{equation}
\Delta_i \equiv x_i - \mu_{\rm s,i}(G_{{\rm RVS},i}),
\end{equation}
where $\mu_{\rm s}(G_{\rm RVS})$ and the corresponding single-star scatter $\sigma_{\rm s}$ are given by the Stage-1 fits. We classify a system as a quadruple candidate if $\Delta_1 > 3\,\sigma_{\rm s,1}$ and $\Delta_2 > 3\,\sigma_{\rm s,2}$. We then define a pool of `single-like' WBs by requiring both components to be consistent with the single-star model, $|\Delta_1|\le \sigma_{\rm s,1}$ and $|\Delta_2|\le \sigma_{\rm s,2}$.

To construct a fair kinematic control sample, we match each quadruple candidate to one system from the single-like pool with similar observational properties. For each WB, we form a feature vector
\begin{equation}
\mathbf{y} = \bigl(d,\, G_{\rm p},\, G_{\rm s},\, \sin\ell,\, \cos\ell,\, b\bigr),
\end{equation}
where $d=1\,000/\varpi$ is the distance using the primary parallax, $(\ell,b)$ are the Galactic coordinates of the primary, and $G_{\rm p}=\min(G_1,G_2)$ and $G_{\rm s}=\max(G_1,G_2)$ are the brighter and fainter component magnitudes. Using $(\sin\ell,\cos\ell)$ avoids the discontinuity at $\ell=0^\circ/360^\circ$. Before matching, each coordinate of $\mathbf{y}$ is centred and scaled using the median and a robust dispersion estimate computed from the control pool, so that no single variable dominates the distance metric. We then select, for each quadruple candidate, the nearest available neighbour in this scaled feature space, with each control system used at most once. This yields a one-to-one matched control sample that closely reproduces the distributions of distance, sky position, and component magnitudes of the quadruple sample.

\subsubsection{Peculiar velocities}

Using \textit{Gaia} astrometry and radial velocities, we compute Galactic space velocities $(U, V, W)$ for the primary component of each system and derive peculiar velocities $|v|$ relative to the Local Standard of Rest $(U, V, W)_{\odot} =(11.1, 12.24, 7.25)~{\rm km\,s^{-1}}$ \citep{Schonrich10}.

\begin{figure}
\includegraphics[width=\columnwidth]{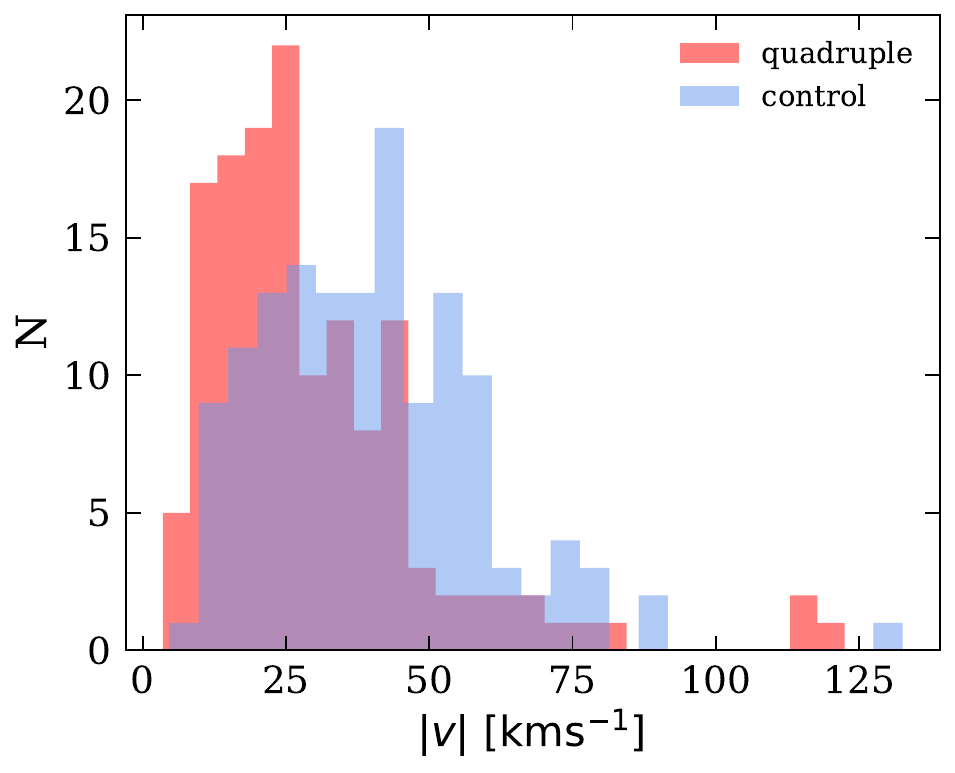}
\caption{Peculiar velocity distributions $|v|$ relative to the Local Standard of Rest for $140$ quadruple systems (red) and a matched control sample of WBs (blue). The control sample is matched in distance, Galactic coordinates $(l,b)$, and primary and secondary magnitudes. Quadruple systems exhibit systematically lower peculiar velocities, indicating a dynamically colder population.
}
\label{fig:v_pec}
\end{figure}

Figure~\ref{fig:v_pec} shows that quadruple systems exhibit systematically lower peculiar velocities with median values of $|v|_{\rm quad}=24.7~{\rm km\,s^{-1}}$ for the quadruples and $|v|_{\rm control}=39.3~ {\rm km\,s^{-1}}$ for the control sample. A Kolmogorov-Smirnov test yields $p =2.32\times 10^{-6}$, indicating a statistically significant difference. 

The systematically lower peculiar velocities of quadruple systems indicate that they belong to a dynamically colder stellar population. Although not explicitly discussed, the velocity dispersions listed in Table 6 of \cite{Tokovinin18} likewise suggest colder kinematics for higher-order multiples relative to triples. Studies of the age–velocity dispersion relation suggest that stars with dispersions comparable to those seen here are predominantly young ($\sim1$ Gyr) \citep[e.g.,][]{Almeida-FernandesRocha-Pinto18, Mustill22, Kontiainen25}. While we do not derive individual stellar ages for our sample, the kinematics of the 2+2 quadruple systems are therefore consistent with a relatively young population, noting that peculiar velocity is only a statistical proxy for age and does not provide individual age estimates. 

\begin{figure}
\includegraphics[width=\columnwidth]{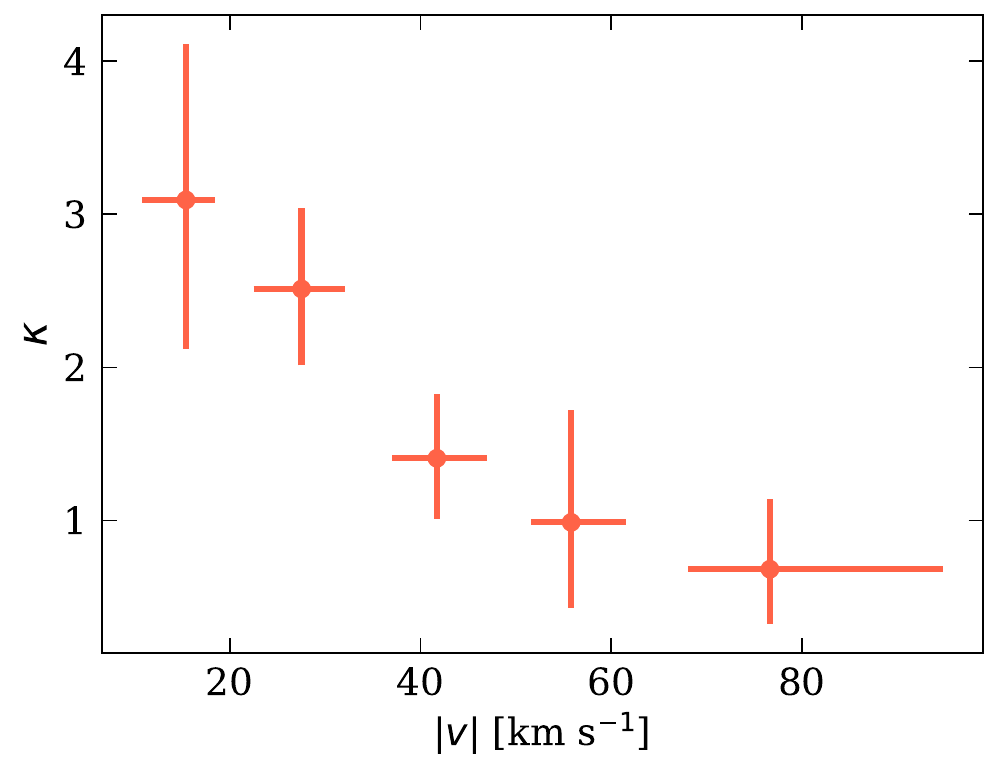}
\caption{Enhancement factor $\kappa = P_{2+2}/p^2$ as a function of peculiar velocity $|v|$ for a subset of stars with -0.2 < [Fe/H] < 0.1 dex. Points show measurements in $|v|$ bins with $1\sigma$ uncertainties from MCMC posteriors. The enhancement declines monotonically with $|v|$, suggesting that the excess of 2+2 systems is associated with dynamically cold, likely younger stellar populations.}
\label{fig:kappa_vpec}
\end{figure}

To explore this connection, we examine in Figure~\ref{fig:kappa_vpec} how the relative incidence of hierarchical configurations changes across bins of peculiar velocity, using the intervals $(0,20,35,50,65,\infty)\,{\rm km\,s^{-1}}$. As in the previous section, to minimise the impact of metallicity-dependent multiplicity
trends, we restrict the analysis to the range -0.2 < [Fe/H] < 0.1 dex. At low peculiar velocities ($|v_{\rm pec}|\lesssim 20~{\rm km\,s^{-1}}$), where the enhancement factor is highest, the WB population includes a substantial fraction of 2+2 systems, at the level of $\sim$8 per cent, alongside a larger fraction of 1+2 ($\sim$16 per cent) configurations. In contrast, at high peculiar velocities ($|v_{\rm pec}|\gtrsim 50~{\rm km\,s^{-1}}$), the fraction of 2+2 systems drops to the $\sim$1 per cent level, while the population becomes increasingly dominated by 1+2 ($\sim$22 per cent) and 1+1 configurations.

This systematic shift occurs despite only modest changes in the overall CB fraction with peculiar velocity, and is therefore driven primarily by a decline in the excess of 2+2 systems. A plausible interpretation is that WBs form with a relatively high fraction of 2+2 configurations, which are progressively depleted over Gyr timescales. A possible outcome is the loss of one of the two inner CBs, for example through merger or dynamical destruction \citep{Kochanek14,HwangZakamska20}, leaving behind a 1+2 hierarchical system. By contrast, evolution into two independent CBs does not appear to be the dominant pathway, since that would be less naturally connected to the observed increase in the 1+2 fraction as the 2+2 fraction declines.


An alternative, non-exclusive explanation is that stars with large peculiar velocities may preferentially include radial migrators from the inner Galaxy, 
where stellar densities and interaction rates are higher \citep[e.g.,][]{Frankel18, Lian22, Zhang25}. If such environments either suppress the long-term survival of fragile hierarchical systems or alter their internal architectures, this could naturally contribute to the observed decline of $\kappa$ with increasing $|v|$. Disentangling age effects from 
migration history will require combining multiplicity statistics with chemical tagging and full Galactic orbit modelling.

\subsubsection{Wide-orbit geometry: the $\theta_{v-r}$ distribution}

The same matched control sample also allows us to test whether 2+2 systems differ in the geometry of their wide-orbit motion. In particular, the angle $\theta_{v-r}$ between the relative position and velocity vectors has been used as a probe of the WB eccentricity distribution \citep{Tokovinin98, Hwang22a}. This diagnostic is especially interesting in light of the unusual orbital properties reported for wide twin binaries \citep{Hwang22b}, toward which our current WB sample is somewhat biased.

We therefore compared the $\theta_{v-r}$ distributions of the 2+2 systems and the matched control sample. Because $\theta_{v-r}$ is an axial circular variable defined on $[0,180^\circ)$, we adopted a two-sample Kuiper test as our primary statistic and found no significant difference between the two samples ($p=0.32$). Thus, within the present sample size, the wide-orbit geometry of 2+2 systems appears broadly similar to that of otherwise comparable WBs. We therefore do not find evidence that the excess of 2+2 systems is accompanied by a markedly different $\theta_{v-r}$ distribution, or equivalently by a strong difference in the WB eccentricity distribution. This null result does not rule out more subtle differences, but it suggests that any imprint of correlated 2+2 formation on the wide-orbit eccentricities is weaker than the clear signal seen in the peculiar velocities.

\section{Conclusions}
\label{sec:conclusions}

Using \textit{Gaia} DR3 RVS data for 9,411 main-sequence WB systems, we have measured the CB fraction $p = 0.150 \pm 0.004$ and the quadruple fraction $P_{2+2} = 0.053 \pm 0.003$.

Our main findings are:
\begin{enumerate}
\item The enhancement factor $\kappa = P_{2+2}/p^2 = 2.34_{-0.11}^{+0.12}$ indicates that quadruple systems are $\sim 2.3$ times more common than expected from independent binary formation.

\item This enhancement is robust and not caused by selection effects, as demonstrated by validation tests: (a) shuffling component pairings while preserving distance and $\Delta G$ distributions; (b) simulations preserving temperature-dependent binary fractions.

\item An independent analysis using PMa (Appendix~\ref{appendix:PMa}) confirms the enhancement, yielding $\kappa_{\rm PMa} = 2.49 \pm 0.65$.

\item The enhancement depends on WB separation. It remains strong at separations $\leq 5\,000$ AU, and shows a decline towards unity beyond $\sim 10\,000$ AU, consistent with a gradual transition to independent pairing.

\item Quadruple systems exhibit systematically lower peculiar velocities compared to matched controls, consistent with a dynamically colder population and, statistically, younger ages.

\item Since $\kappa$ and $p$ uniquely determine the partition of close pairs among binaries, triples, and quadruples, the observed decline of $\kappa$ with increasing peculiar velocity implies a relative conversion of 2+2 quadruples into triples in dynamically hotter populations.

\end{enumerate}

\section*{Acknowledgements}
We thank the anonymous referee for a constructive report. We thank the \textit{Gaia} Consortium, and in particular the CU6 team, for their outstanding efforts in the development and validation of the \textit{Gaia} RVS pipeline, without which this work would not have been possible. We thank Andrei Tokovinin for valuable comments and suggestions on an early draft of this manuscript. We are grateful to Shion Andrew for sharing her implementation and calibration of the \textit{Gaia} DR2–DR3 PMa frame-rotation correction used in our Appendix~\ref{appendix:PMa} consistency check. 
DB acknowledges the support of the Blavatnik family and the British Friends of the Hebrew University (BFHU) and Didier Queloz for his courteous hospitality.

\section*{Data Availability}

The full list of sources used in this analysis as well as the sub-sample of quadruple candidates is provided in the online supplementary material. The \textit{Gaia} DR3 data are publicly available at \url{https://gea.esac.esa.int/archive/}. The WB catalogue of \citet{El-Badry21} is available at \url{https://zenodo.org/records/4435257}. The code and derived data products used in this paper will be made available upon reasonable request to the corresponding author.



\bibliographystyle{mnras}
\bibliography{Refs}



\appendix
\section{Proper Motion Anomaly Confirmation of Quadruple Excess}
\label{appendix:PMa}

As an independent check, we use the proper-motion anomaly (PMa) method to flag unresolved companions via astrometric accelerations. We define the PMa vector
\begin{equation}
\Delta\boldsymbol{\mu} \equiv \boldsymbol{\mu}_{\rm DR3} - \boldsymbol{\mu}_{{\rm DR2}\rightarrow{\rm DR3}},
\end{equation}
where $\boldsymbol{\mu}_{{\rm DR2}\rightarrow{\rm DR3}}$ denotes the \textit{Gaia} DR2 proper motion transformed into the DR3 reference frame by applying a small, magnitude-dependent frame-rotation correction. Following the approach in \cite{Lindegren20} and the calibration procedure described in Andrew (thesis, unpublished), the rotation vector $\boldsymbol{\omega}(G)$ is estimated in 0.5-mag bins and used to correct the DR2 proper motions prior to computing $\Delta\boldsymbol{\mu}$.

We define a PMa significance
\begin{equation}
S_{\rm PMa} \equiv \frac{|\Delta\boldsymbol{\mu}|}{\sigma_{\Delta\mu}},
\end{equation}
where $\sigma_{\Delta\mu}$ is computed from the quadrature sum of the DR2 and DR3 proper-motion uncertainties. We classify sources with $S_{\rm PMa} > 5$ as binary candidates, a conservative threshold calibrated to yield a negligible contamination rate from single stars in simulated \textit{Gaia}-like astrometric solutions \citep{Belokurov2020,Penoyre20}.

PMa is complementary to the RVS-based RV-variability selection in the main text because it is most sensitive to companions with orbital timescales of years to decades that induce sky-plane accelerations, whereas RV variability preferentially detects shorter-period systems with large line-of-sight velocities. PMa sensitivity depends on brightness and distance and may be affected by residual systematics. We therefore use it as a consistency check rather than as our primary CB classifier.

Using the same RVS-selected WB sample as in the main text to avoid changes in the parent selection, we measure the pooled PMa binary fraction $p \equiv N_{\rm bin}/(2N_{\rm pair})$ and the fraction of pairs where both components are flagged, $P_{2+2} \equiv N_{12}/N_{\rm pair}$, and compute the enhancement factor $\kappa_{\rm PMa}=P_{2+2}/p^2$. We obtain
\begin{equation}
\kappa_{\rm PMa} = 2.49 \pm 0.65,
\end{equation}
consistent with the RVS-based value and supporting the reality of the quadruple excess.

\bsp	
\label{lastpage}
\end{document}